\documentclass[useAMS,usenatbib]{mn2e}
\usepackage{graphicx,txfonts}

\defcitealias{ChandH_04a}{C04}
\defcitealias{LevshakovS_06b}{L06}
\defcitealias{LevshakovS_07a}{L07}
\defcitealias{MurphyM_03a}{M03}
\defcitealias{MurphyM_04a}{M04}

\newcommand{\kms}{\hbox{${\rm km\,s}^{-1}$}}
\newcommand{\zem}{\hbox{$z_{\rm em}$}}
\newcommand{\zab}{\hbox{$z_{\rm abs}$}}

\newcommand{\SNR}{\hbox{${\rm S/N}$}}
\newcommand{\da}{\hbox{$\Delta\alpha/\alpha$}}

\newcommand{\bspsmall}{\vspace{0.5cm}\small\noindent This paper has been typeset
from a \TeX/\LaTeX\ file prepared by the author.\normalsize}

\title[VLT/UVES constraints on varying $\alpha$]{Revision of VLT/UVES constraints on a varying fine-structure constant}

\author[M. T. Murphy, J. K. Webb, V. V. Flambaum]{M. T. Murphy,$^{1,2}$\thanks{E-mail: mmurphy@swin.edu.au (MTM)}, J. K. Webb,$^{3}$ V. V. Flambaum$^{3}$\\
$^{1}$Centre for Astrophysics and Supercomputing, Swinburne University of Technology, Mail H39, PO Box 218, Victoria 3122, Australia\\
$^{2}$Institute of Astronomy, University of Cambridge, Madingley Road,
  Cambridge, CB3 0HA, UK\\
$^{3}$School of Physics, University of New South Wales, UNSW Sydney
  N.S.W. 2052, Australia}

\begin{document}

\date{Accepted ---. Received ---; in original form ---}

\pagerange{\pageref{firstpage}--\pageref{lastpage}}

\pubyear{2007}

\maketitle

\label{firstpage}

\begin{abstract}
  We critically review the current null results on a varying
  fine-structure constant, $\alpha$, derived from VLT/UVES quasar
  absorption spectra, focusing primarily on the many-multiplet
  analysis of 23 absorbers from which \citet{ChandH_04a} reported a
  weighted mean relative variation of
  $\da=(-0.06\pm0.06)\times10^{-5}$. Our analysis of the \emph{same
    reduced data}, using the \emph{same fits to the absorption
    profiles}, yields very different individual $\da$ values with
  uncertainties typically larger by a factor of $\sim$3. We attribute
  the discrepancies to flawed parameter estimation techniques in the
  original analysis and demonstrate that the original $\da$ values
  were strongly biased towards zero. Were those flaws not present, the
  input data and spectra should have given a weighted mean of
  $\da=(-0.44\pm0.16)\times10^{-5}$. Although this new value
  \emph{does} reflect the input \emph{spectra and fits} (unchanged
  from the original work -- only our analysis is different), we do not
  claim that it supports previous Keck/HIRES evidence for a varying
  $\alpha$: there remains significant scatter in the individual $\da$
  values which may stem from the overly simplistic profile fits in the
  original work. Allowing for such additional, unknown random errors
  by increasing the uncertainties on $\da$ to match the scatter
  provides a more conservative weighted mean,
  $\da=(-0.64\pm0.36)\times10^{-5}$. We highlight similar problems in
  other current UVES constraints on varying $\alpha$ and argue that
  comparison with previous Keck/HIRES results is premature.
\end{abstract}


\section{Introduction}\label{sec:intro}

Absorption lines from heavy element species in distant gas clouds
along the sight-lines to background quasars (QSOs) are important
probes of possible variations in the fine-structure constant,
$\alpha$, over cosmological time- and distance-scales. The
many-multiplet (MM) method \citep{DzubaV_99a,WebbJ_99a} utilizes the
relative wavelength shifts expected from different transitions in
different neutral and/or ionized metallic species to measure $\alpha$
with an order of magnitude better precision than previous techniques
such as the alkali doublet (AD) method. It yielded the first tentative
evidence for $\alpha$-variation \citep{WebbJ_99a} and subsequent,
larger samples saw this evidence grow in significance and internal
robustness (\citealt{MurphyM_01a}; \citealt{WebbJ_01a};
\citealt*[hereafter \citetalias{MurphyM_03a}]{MurphyM_03a}).  MM
analysis of 143 absorption spectra, all from the Keck/HIRES
instrument, currently indicate a smaller $\alpha$ in the clouds at the
fractional level $\da=(-0.57\pm0.11)\times10^{-5}$ over the redshift
range $0.2<\zab<4.2$ \citep[][hereafter
\citetalias{MurphyM_04a}]{MurphyM_04a}. Clearly, this potentially
fundamental result must be refuted or confirmed with many different
spectrographs to guard against subtle systematic errors which, despite
extensive searches (\citealt{MurphyM_01b}; \citetalias{MurphyM_03a}),
have evaded detection so far.

\begin{figure*}
\centerline{\includegraphics[width=0.95\textwidth]{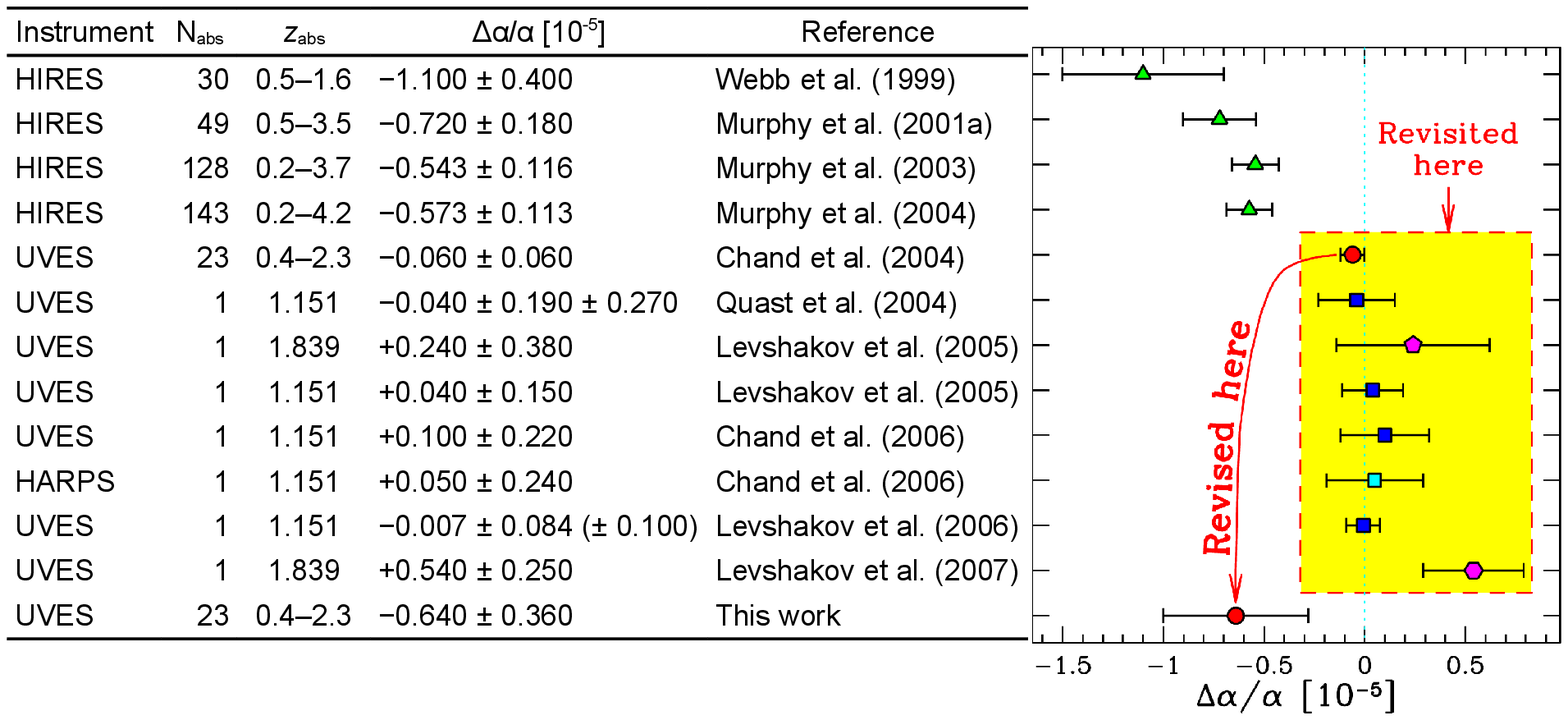}}
\vspace{-2mm}
\caption{Summary of the many-multiplet constraints on $\da$ in the
  literature. The table shows the instrument used, the sample size
  (N$_{\rm abs}$), the absorption redshift (or range), the value of
  $\da$ with 1-$\sigma$ error (for statistical samples, the weighted
  mean value and its 1-$\sigma$ error are quoted) and the
  corresponding references. The plot allows easy comparison of the
  $\da$ values and their 1-$\sigma$ errors but there are several
  important caveats the reader should note: (i) We demonstrate in this
  paper that some UVES and HARPS constraints are based on flawed
  parameter estimation techniques and are not reliable. The
  reliability of other UVES constraints is also questioned in this
  work; (ii) Many points are not independent.  Points in the plot with
  the same symbols use the same data (to varying extents); (iii) The
  sample sizes are very different.  Some samples contain only a single
  absorption system; random errors for statistical samples are
  systematic errors for single absorbers; (iv) The redshift ranges
  vary considerably. If $\alpha$ does vary from absorber to absorber
  then that information is lost here; (v) The typical signal-to-noise
  ratio of the spectra in each sample is different; (vi) Some
  error-bars on $\da$ contain a systematic component which is combined
  with the statistical one into a single error term. Other points
  contain no systematic error term.  \citet{QuastR_04a} quote a
  separate systematic term.  \citet{LevshakovS_06b} do calculate a
  systematic term but do not quote it together with the statistical
  term.}
\label{fig:summary}
\end{figure*}

First attempts at constraining $\alpha$-variation with the Ultraviolet
and Visual Echelle Spectrograph (UVES) on the Very Large Telescope
(VLT) in Chile have, at first glance, yielded null results. The MM
constraints in the literature are summarized by
Fig.~\ref{fig:summary}, the caption of which describes several
important caveats for interpreting the figure.

To date, the only MM analysis of a statistically significant UVES
sample is that of \citet[hereafter \citetalias{ChandH_04a};
\citealt{SrianandR_04a} summarize the main results]{ChandH_04a} who
reported that $\da=(-0.06\pm0.06)\times10^{-5}$ from 23 absorbers in
the redshift range $0.4<\zab<2.3$. The main aim of the current paper
is to revise these results after demonstrating simple flaws in the
data analysis technique of \citetalias{ChandH_04a}. The same flaws are
also evident in the AD analysis of \citet{ChandH_05a} who reported a
weighted mean $\da=(+0.15\pm0.43)\times10^{-5}$ from 15 Si{\sc \,iv}
doublets over the range $1.5<\zab<3.0$.

Different MM analyses of 2 individual absorption clouds for which
higher signal-to-noise ratio (\SNR) spectra are available have also
provided seemingly strong constraints on $\da$. Null results with
1-$\sigma$ statistical uncertainties of $0.19$, $0.15$ and
$0.24\times10^{-5}$ were derived by \citet{QuastR_04a},
\citet{LevshakovS_05a} and \citet{ChandH_06a}, respectively, from UVES
spectra of the complex $\zab=1.151$ absorber towards HE\,0515$-$4414.
The constraint of \citet{ChandH_06a} again suffers from the same data
analysis errors as the statistical results in \citetalias{ChandH_04a}.
\citet[hereafter \citetalias{LevshakovS_06b}]{LevshakovS_06b} improved
their earlier constraints (from \citealt{QuastR_04a} and
\citealt{LevshakovS_05a}) to $\da=(-0.007\pm0.084)\times10^{-5}$. We
demonstrate here that the UVES data utilized in that analysis simply
do not allow such a low statistical uncertainty. The other individual
high-\SNR\ absorber analysed with the MM method is at $\zab=1.839$
towards Q\,1101$-$264. Recently, \citet[hereafter
\citetalias{LevshakovS_07a}]{LevshakovS_07a} revised their earlier
constraint \citep{LevshakovS_05a} from
$\da=(0.20\pm0.38)\times10^{-5}$ to $\da=(0.54\pm0.25)\times10^{-5}$
by analysing new spectra with higher resolution. We discuss this
result further in Section \ref{sec:other}.

As mentioned above, the main focus of this paper is to critically
analyse the reliability of the \citetalias{ChandH_04a} results, the
only statistical MM study apart from our previous Keck/HIRES work. In
Section \ref{sec:motiv} we point out problems in the `$\chi^2$-curve'
measurement technique used by \citetalias{ChandH_04a}. We also
introduce a simple algorithm for estimating the minimum statistical
error in $\da$ achievable from any given absorption spectrum. The
uncertainties quoted by \citetalias{ChandH_04a} are inconsistent with
this `limiting precision'. In Section \ref{sec:rev} we revise the
\citetalias{ChandH_04a} results using the same spectral data but with
robust numerical algorithms which we demonstrate are immune to the
errors evident in \citetalias{ChandH_04a}. Section \ref{sec:other}
discusses the other constraints on $\da$ from UVES mentioned above. We
conclude in Section \ref{sec:conc}.

\section{Motivations for revising UVES results}\label{sec:motiv}

\subsection{$\chi^2$ curves}\label{ssec:chi}

$\da$ is typically measured in a quasar absorption system using a
$\chi^2$ minimization analysis of multiple-component Voigt profiles
simultaneously fit to the absorption profiles of several different
transitions. The column densities, Doppler widths and redshifts
defining the individual components are varied iteratively until the
decrease in $\chi^2$ between iterations falls below a specified
tolerance, $\Delta\chi^2_{\rm tol}$. Our approach in
\citet{MurphyM_01a}, \citetalias{MurphyM_03a} \&
\citetalias{MurphyM_04a} was simply to add $\da$ as an additional
fitting parameter, to be varied simultaneously with all the other
parameters in order to minimize $\chi^2$. The approach of
\citetalias{ChandH_04a}, following \citet{WebbJ_99a}, was to keep
$\da$ as an external parameter: for a fixed input value of $\da$ the
other parameters of the fit are varied to minimize $\chi^2$. The input
value of $\da$ is stepped along over a given range around zero and
$\chi^2$ is computed at each step. The functional form of $\chi^2$
implies that, in the vicinity of the best-fitting $\da$, the `$\chi^2$
curve' -- the value of $\chi^2$ as a function of the input value of
$\da$ -- should be near parabolic and smooth. In practice, this means
that in each separate fit, with a different input $\da$,
$\Delta\chi^2_{\rm tol}$ should be set to $\ll 1$ to ensure that any
fluctuations on the final $\chi^2$ curve are also $\ll 1$. This is
obviously crucial when using the standard method of deriving the
1-$\sigma$ uncertainty, $\delta(\da)$, from the width of the $\chi^2$
curve at $\chi^2_{\rm min}+1$: if the fluctuations on the $\chi^2$
curve are $\ga$$1$ then one expects $\chi^2_{\rm min}$ to be rather
poorly defined and $\delta(\da)$ to be underestimated. The larger the
fluctuations on the $\chi^2$ curve, the more the measured value of
$\da$ will deviate from the true value and, even worse, the more
\emph{significantly} it will deviate since its uncertainty will be
more underestimated.

Even a cursory glance at the $\chi^2$ curves of
\citetalias{ChandH_04a} -- figure 2 in \cite{SrianandR_04a}, figure 14
in \citetalias{ChandH_04a} itself -- reveal that \emph{none} could be
considered smooth at the $\ll 1$ level and, almost without exception,
the $\chi^2$ fluctuations significantly exceed unity (two examples are
shown in Section \ref{ssec:bias} -- see Fig.~\ref{fig:chi}). Again, we
stress that no matter how noisy the spectral data or how poorly one's
model profile fits the data or how many free parameters are being
fitted\footnote{Of course, the number of parameters fitted must be
  less than the number of spectral pixels.}, the $\chi^2$ curve should
be smooth and near parabolic in the vicinity of the best fit. The
$\chi^2$ fluctuations in \citetalias{ChandH_04a} must therefore be due
to failings in the algorithm used to minimize $\chi^2$ for each input
$\da$. This point can not be over-emphasized since the $\chi^2$ curve
is the very means by which \citetalias{ChandH_04a} measure $\da$ and
its uncertainty in each absorption system.

Based on fits to simulated absorption spectra, \citetalias{ChandH_04a}
argue that their measurement technique is indeed robust. However,
strong fluctuations even appear in the $\chi^2$ curves for these
simulations (their figure 2). This leads to spurious $\da$ values:
figure 6 in \citetalias{ChandH_04a} shows the results from 30
realizations of a simulated single-component Mg/Fe{\sc \,ii} absorber.
At least 15 $\da$ values deviate by $\ge1\sigma$ from the input value;
8 of these deviate by $\ge2\sigma$ and 4 by $\ge3\sigma$. There is
even a $\approx5$-$\sigma$ value. The distribution of $\da$ values
should be Gaussian in this case but these outliers demonstrate that it
obviously is not. The $\chi^2$ fluctuations also cause the uncertainty
estimates, $\delta(\da)$, from each simulation to range over a factor
of $\approx4$ even though all realizations had the same simulated
\SNR\ and input parameters. None of these problems arise in our own
simulations of either single- or multiple-component systems
(\citealt{MurphyM_02b}; \citetalias{MurphyM_03a}).

Clearly, the results of \citetalias{ChandH_04a} cannot be reliable if
the $\chi^2$ minimization algorithm -- the means by which $\da$ and
$\delta(\da)$ are measured -- failed. From the discussion above, we
should expect that their uncertainty estimates are underestimated as a
result. The following sub-section demonstrates this by introducing a
simple measure of the minimum possible $\delta(\da)$ in a given
absorption system. We correct the analysis of \citetalias{ChandH_04a}
in Section \ref{sec:rev} using the same data and profile fits.

\subsection{A simple measure of the limiting precision on
  $\da$}\label{ssec:limit}

\subsubsection{Formalism}

\begin{figure*}
\centerline{\hbox{
 \includegraphics[width=0.45\textwidth]{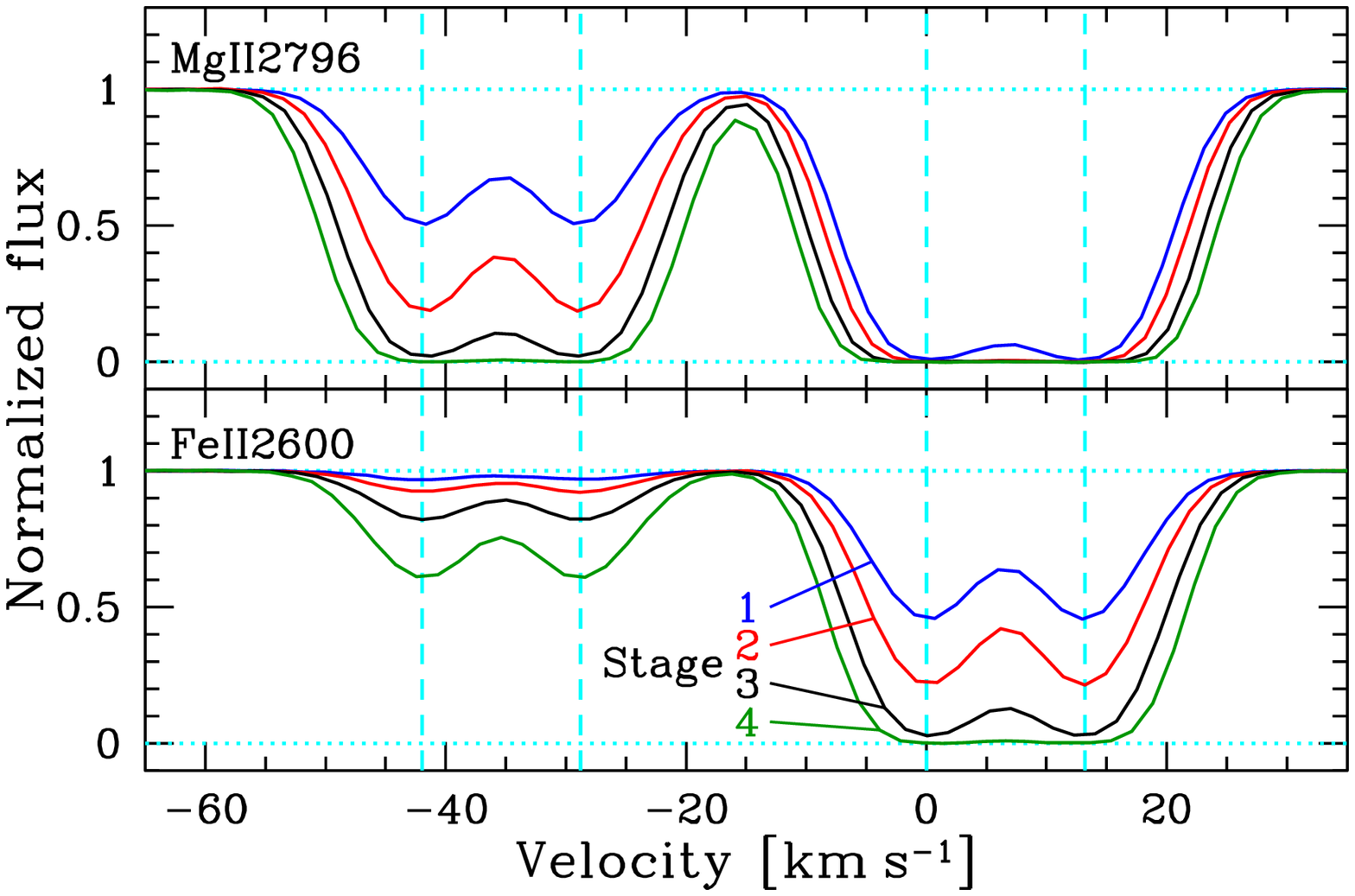}
 \hspace{0.01\textwidth}
 \includegraphics[width=0.466\textwidth]{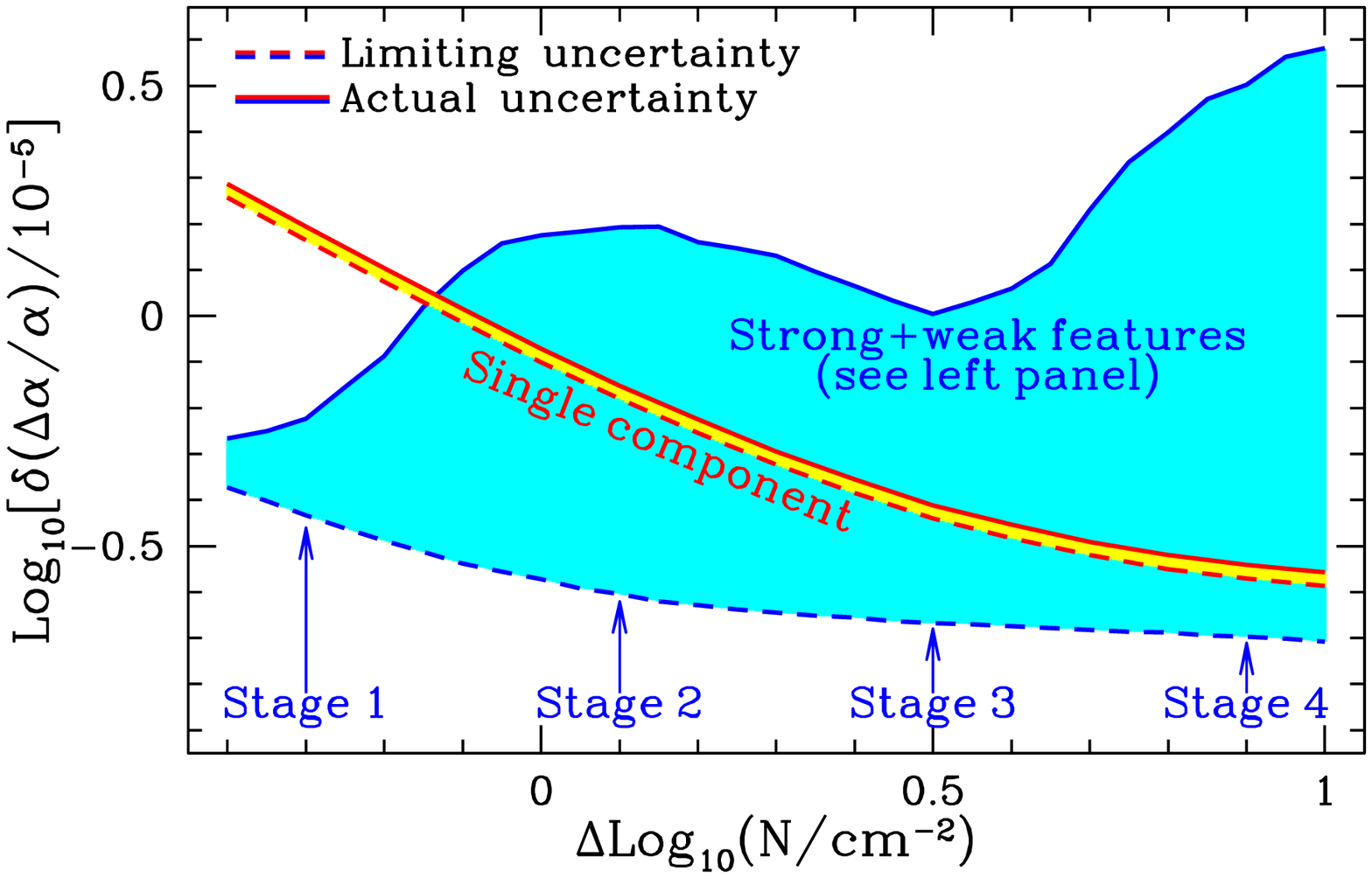}
}}
\vspace{-2mm}
\caption{Left-hand panel: Simulation of two transitions in a
  multi-component absorber. Both transitions have the same velocity
  structure: two main spectral features, each comprising two blended
  velocity components with equal column densities. Labelled are
  distinct stages of differential saturation in the two main spectral
  features. Right-hand panel: The difference between the limiting
  precision, $\delta(\da)_{\rm lim}$, and the actual precision (as
  derived by $\chi^2$-minimization analysis) varies strongly from
  stage to stage. In a single-component absorber the actual
  uncertainty tracks $\delta(\da)_{\rm lim}$, as expected. See text
  for discussion.}
\label{fig:sim}
\end{figure*}

The velocity shift, $\Delta v_i$, of transition $i$ due to a small
variation in $\alpha$, i.e.~$\da\ll 1$, is determined by the
$q$-coefficient for that transition,
\begin{equation}\label{eq:da}
\omega_{i}(z) \equiv \omega_{i}(0) + q_i\left[\left(\alpha_z/\alpha_0\right)^2-1\right]\,\hspace{1em}\Rightarrow\hspace{1em}\frac{\Delta v_i}{c} \approx -2\frac{\Delta\alpha}{\alpha}\frac{q_i}{\omega_{i}(0)}\,,
\end{equation}
where $\omega_{i}(0)$ and $\omega_{i}(z)$ are the rest-frequencies in
the laboratory and in an absorber at redshift $z$ respectively.
Similarly, $\alpha_0$ and $\alpha_z$ are the laboratory and absorber
values of $\alpha$. The MM method is the comparison of measured
velocity shifts from several transitions (with different
$q$-coefficients) to compute the best-fitting $\da$. The linear equation
(\ref{eq:da}) implies that the error in $\da$ is determined only by
the distribution of $q$-coefficients (assumed to have negligible
errors) being used and the statistical errors in the velocity shifts,
$\delta_{{\rm v},i}$:
\begin{equation}\label{eq:lim1}
\delta(\da)_{\rm lim} = \sqrt{S/D}\,,
\end{equation}
where
\begin{equation}\label{eq:lim2}
S\equiv\sum_i\left(\frac{\delta_{{\rm v},i}}{c}\right)^{-2}
\end{equation}
and
\begin{equation}
D\equiv S\sum_i\left(\frac{2q_i}{\omega_{i}(0)}\right)^2\left(\frac{\delta_{{\rm v},i}}{c}\right)^{-2} - \left[\sum_i\frac{2q_i}{\omega_{i}(0)}\left(\frac{\delta_{{\rm v},i}}{c}\right)^{-2}\right]^2\,.
\end{equation}
This expression is just the solution to a straight-line least-squares
fit, $y=a+bx$, to data $(x_j,y_j)$, with errors only on the $y_j$,
where the intercept $a$ is also allowed to vary. Allowing the
intercept to vary is important since it mimics the real situation in
fitting absorption lines where the absorption redshift and $\da$ must
be determined simultaneously.

Equation (\ref{eq:lim1}) can only be used if one knows the statistical
error on the velocity shift measurement for each transition,
$\delta_{{\rm v},i}$. This quantity is only well-defined in an
absorption system with a single fitted velocity component or in a
system with several velocity components which do not blend or overlap
significantly with each other. However, the general case -- and,
observationally, by far the most common one -- is that absorbers have
many velocity components which, at the resolution of the spectrograph,
are strongly blended together. For this general case we wish to define
a `total velocity uncertainty' for each transition \emph{integrated
  over the absorption profile} (i.e.~over all components),
$\sigma_{{\rm v},i}$, so that the substitution $\delta_{{\rm
    v},i}=\sigma_{{\rm v},i}$ in equation (\ref{eq:lim1}) provides
some easily-interpreted information about $\delta(\da)_{\rm lim}$.

The quantity $\sigma_{{\rm v},i}$ is commonly used in radial-velocity
searches for extra-solar planets, e.g.~\citet{BouchyF_01a}, but is not
normally useful in QSO absorption-line studies. Most metal-line QSO
absorption profiles display a complicated velocity structure and one
usually focuses on the properties of individual velocity components,
each of which is typically modelled by a Voigt profile. However, it is
important to realize that $\da$ and its uncertainty are integrated
quantities determined by the entire absorption profile. Some velocity
components -- typically the narrow, deep-but-unsaturated, isolated
ones -- will obviously provide stronger constraints than others but
all components nevertheless contribute something. Thus, $\sigma_{{\rm
    v},i}$ should incorporate all the velocity-centroiding information
available from a given profile shape. From a spectrum $F(k)$ with
1-$\sigma$ error array $\sigma_F(k)$, the minimum possible velocity
uncertainty contributed by pixel $k$ is given by \citep{BouchyF_01a}
\begin{equation}\label{eq:sv_i}
\frac{\sigma_{\rm v}(k)}{c} = \frac{\sigma_F(k)}{\lambda(k)\,\left[\partial F(k)/\partial\lambda(k)\right]}\,.
\end{equation}
That is, a more precise velocity measurement is available from those
pixels where the flux has a large gradient and/or a small uncertainty.
This quantity can be used as an optimal weight, $W(k) \equiv
\left[\sigma_{\rm v}(k)/c\right]^{-2}$, to derive the total velocity
precision available from all pixels in a portion of spectrum,
\begin{equation}\label{eq:sv}
\textstyle\sigma_{\rm v} = c\left[\sum_k W(k)\right]^{-1/2}\,.
\end{equation}

For each transition in an absorber, $\sigma_{{\rm v},i}$ is calculated
from equations (\ref{eq:sv_i}) \& (\ref{eq:sv}). Note that the only
requirements are the 1-$\sigma$ error spectrum and the multi-component
Voigt profile fit to the transition's absorption profile. The latter
allows the derivative in (\ref{eq:sv_i}) to be calculated without the
influence of noise. If very high \SNR\ spectra are available --
i.e.~where the $\sigma_F(k)$ are always much less than the flux
difference between neighbouring pixels in high-gradient portions of
the absorption profiles -- then one could use the spectrum itself
instead of the Voigt profile fit, thus making the estimate of
$\sigma_{\rm v}$ model-independent. Once $\sigma_{{\rm v},i}$ has been
calculated for all transitions $i$, the uncertainty in $\da$ simply
follows from equation (\ref{eq:lim1}) with the substitution
$\delta_{{\rm v},i}=\sigma_{{\rm v},i}$.

\subsubsection{Limiting precision}

It is important to realize that the uncertainty calculated with the
above method represents the absolute minimum possible 1-$\sigma$
error on $\da$; the real error -- as derived from a simultaneous
$\chi^2$-minimization of all parameters comprising the Voigt profile
fits to all transitions -- will always be larger than
$\delta(\da)_{\rm lim}$ from equation (\ref{eq:lim1}). The main reason
for this is that absorption systems usually have several velocity
components which have different optical depths in different
transitions. Equation (\ref{eq:lim1}) assumes that the velocity
information integrated over all components in one transition can be
combined with the same integrated quantity from another transition to
yield an uncertainty on $\da$. However, in a real determination of
$\da$, each velocity component (or group of components which define a
sharp spectral feature) in one transition is, effectively, compared
with only the same component (or group) in another transition.

Figure~\ref{fig:sim} illustrates this important point. It shows
simulated absorption profiles for two transitions commonly used in MM
analyses (Mg{\sc \,ii} $\lambda$2796 and Fe{\sc \,ii} $\lambda$2600)
in different stages of saturation. The velocity structure is identical
for both transitions and contains two well-separated main spectral
features (MSFs), each of which comprises two velocity components which
are blended together. The column-density ratios between the
corresponding components of the two transitions is kept fixed while
the total column density is varied. For each simulation with a
different total column density we determined $\delta(\da)_{\rm lim}$
using the method above and the real value of $\delta(\da)$ using the
usual $\chi^2$ minimization analysis. When comparing $\delta(\da)_{\rm
  lim}$ with $\delta(\da)$ one notices 4 characteristic stages as the
column density increases:
\begin{itemize}
\item {\bf Stage 1:} Both MSFs are relatively unsaturated in both
  transitions and so $\sigma_{{\rm v},i}$ will be small for $i$=Mg{\sc
    \,ii} $\lambda$2796 and $i$=Fe{\sc \,ii} $\lambda$2600. Thus,
  $\delta(\da)_{\rm lim}$ is quite small. However, note that the
  right-hand MSF in Mg{\sc \,ii} $\lambda$2796 is nevertheless a
  little saturated and so the \emph{real} precision is somewhat
  weakened, i.e.~$\delta(\da)$ is pushed higher than $\delta(\da)_{\rm
    lim}$; the high velocity precision available from that MSF in
  Fe{\sc \,ii} $\lambda$2600 is `wasted' because the profile of the
  \emph{corresponding} MSF in Mg{\sc \,ii} $\lambda$2796 is smoother.
\item {\bf Stage 2:} The right-hand MSF in Mg{\sc \,ii} $\lambda$2796
  is now completely saturated. Since that part of the profile is now
  smoother, $\delta(\da)_{\rm lim}$ should get \emph{larger}. However,
  this is more than compensated by the additional centroiding
  potential (or velocity information) now offered by the weaker
  velocity components in the left-hand MSF \emph{and} both MSFs in
  Fe{\sc \,ii} $\lambda$2600 due to the increased column density. On
  the other hand, the \emph{real} precision, $\delta(\da)$, has
  substantially worsened because the right-hand MSF from the two
  transitions no longer constrain $\da$ tightly when considered
  together. This principle also applies to the left-hand MSF where, in
  Fe{\sc \,ii} $\lambda$2600, it is too weak to provide strong
  constraints, even though the same velocity components in Mg{\sc
    \,ii} $\lambda$2796 are stronger and well-defined.
\item {\bf Stage 3:} The decrease in $\delta(\da)_{\rm lim}$ is now
  dominated by the small increase in velocity information available
  from the left-hand MSF because the right-hand MSF of both
  transitions is now saturated. Note also that the \emph{real}
  precision also improves here because the components of the left-hand
  MSF in Fe{\sc \,ii} $\lambda$2600 are getting stronger while the
  corresponding components of Mg{\sc \,ii} $\lambda$2796 are not
  completely saturated.
\item {\bf Stage 4:} The decrease in $\delta(\da)_{\rm lim}$ is now
  only marginal because it is dominated only by one MSF in one
  transition, i.e.~the left-hand side of Fe{\sc \,ii} $\lambda$2600.
  However, $\delta(\da)$ has increased sharply because now
  even the left-hand MSF of Mg{\sc \,ii} $\lambda$2796 is saturated
  and, when considered together with the corresponding MSF of Fe{\sc
    \,ii} $\lambda$2600, provides no constraint on $\da$.
\end{itemize}
To summarize this illustration, it is always the case that
$\delta(\da)_{\rm lim}<\delta(\da)$ and it is the degree of
\emph{differential saturation} between \emph{corresponding} components
of different transitions which determines how much worse $\delta(\da)$
is than $\delta(\da)_{\rm lim}$. Also note that the \SNR\ of the data
(or the simulations above) affects $\delta(\da)_{\rm lim}$ and
$\delta(\da)$ in precisely the same way. That is, the ratio
$\delta(\da)_{\rm lim}/\delta(\da)$ is independent of \SNR. Only the
degree of differential saturation between corresponding components of
different transitions affects $\delta(\da)_{\rm lim}/\delta(\da)$.

Finally, as the above discussion implies, fits comprising a single
velocity component (or multiple but well separated components) should
have $\delta(\da)_{\rm lim}\approx\delta(\da)$. As an internal
consistency check on our simulations, Fig.~\ref{fig:sim} also shows
the results for an absorber with a single velocity component, again in
the Mg{\sc \,ii} $\lambda$2796 and Fe{\sc \,ii} $\lambda$2600
transitions. Note that $\delta(\da)_{\rm lim}$ tracks quite closely
the real value of $\delta(\da)$ as a function of column density.
Nevertheless, the real error is slightly worse than $\delta(\da)_{\rm
  lim}$; this is expected because the real estimate derives from a fit
where all parameters of the absorption profiles, including the Doppler
parameters and column densities, are varied simultaneously, thus slightly
weakening the constraint on $\da$ (and the absorber redshift).

\subsubsection{Application to existing constraints on $\da$}\label{sssec:apply}

We have calculated $\delta(\da)_{\rm lim}$ for the absorbers from the
three independent data-sets which constitute the strongest current
constraints on $\Delta\alpha/\alpha$: (i) the 143 absorbers in our
Keck/HIRES sample \citepalias{MurphyM_04a}; (ii) the 23 absorption
systems, comprising mostly Mg/Fe{\sc \,ii} transitions, from UVES
studied by \citetalias{ChandH_04a}; (iii) the UVES exposures of the
$z_{\rm abs}=1.151$ absorber towards HE\,0515$-$4414 studied by
\citetalias{LevshakovS_06b}. Calculating $\delta(\da)_{\rm lim}$
requires only the error spectra and the Voigt profile models used to
fit the data. For samples (ii) \& (iii) we use the Voigt profile
models published by those authors. Below we describe the errors
arrays. For sample (ii), other aspects of the data are important for
the analysis in Section \ref{sec:rev} and so we also describe them
here.

The reduced (i.e.~one-dimensional) spectra in sample (ii) were kindly
provided to us by B.~Aracil who confirmed that the wavelength and flux
arrays are identical to those used in \citetalias{ChandH_04a}.
However, one main difference is that the error arrays we use are
generally a factor $\approx$1.4 smaller than those used by
\citetalias{ChandH_04a} (H.~Chand, B.~Aracil, 2006, private
communication). We have confirmed this by digitizing the absorption
profiles plotted in \citetalias{ChandH_04a}. The reason for this is
that they derive their error arrays by adding two error terms of
similar magnitude in quadrature, even though each term should
reasonably approximate the actual error. One term reflects the formal
photon statistics while the other reflects the r.m.s.~variation in the
flux from the different exposures which are combined to form the final
spectrum. Our error spectra were derived from the maximum of the two
terms. Thus, the error spectra of \citetalias{ChandH_04a} are
$\approx$1.3--2 times larger than ours. We have confirmed that our
error arrays match well the r.m.s.~flux in unabsorbed spectral
regions; they therefore more accurately reflect the real uncertainty
in flux for each spectral pixel. Note that this implies that
$\delta(\da)_{\rm lim}$ calculated using our spectra will be
\emph{smaller} than the value \citetalias{ChandH_04a} would derive.
The only other difference between our spectra and those of
\citetalias{ChandH_04a} is that we performed our own continuum
normalization of the absorption profiles.  However, we used a method
similar to that employed by \citetalias{ChandH_04a} and any small
differences will have negligible effects on the analysis here and in
Section \ref{sec:rev}\footnote{This was checked by simply fitting
  different continua to some spectra and observing the effect on the
  best-fitting $\da$ for the absorbers involved.}.

For sample (iii), we reduced the raw UVES exposures using a modified
version of the UVES pipeline. For the present analysis, small
differences between our reduction and that of \citetalias{LevshakovS_06b}
are unimportant; all that is required is that the error arrays
match fairly closely. Indeed, the \SNR\ matches very well those quoted
by \citetalias{LevshakovS_06b} in the relevant portions of the reduced
spectrum.

For all samples, the atomic data for the transitions (including
$q$-coefficients) were the same as used by the original authors.

In practice, when applying equations (\ref{eq:sv_i}) and (\ref{eq:sv})
we sub-divide the absorption profile of each transition into $15{\rm
  \,km\,s}^{-1}$ chunks to mitigate the effects illustrated in
Fig.~\ref{fig:sim}. This provides a value of $\delta(\da)_{{\rm
    lim},j}$ for each chunk $j$. The final value of $\delta(\da)_{\rm
  lim}$ is simply $\{\sum_j 1/[\delta(\da)_{{\rm lim},j}]^2\}^{1/2}$;
in all cases this is $<$1.4 times the value obtained without
sub-divisions.

\begin{figure}
\centering
\includegraphics[width=\columnwidth]{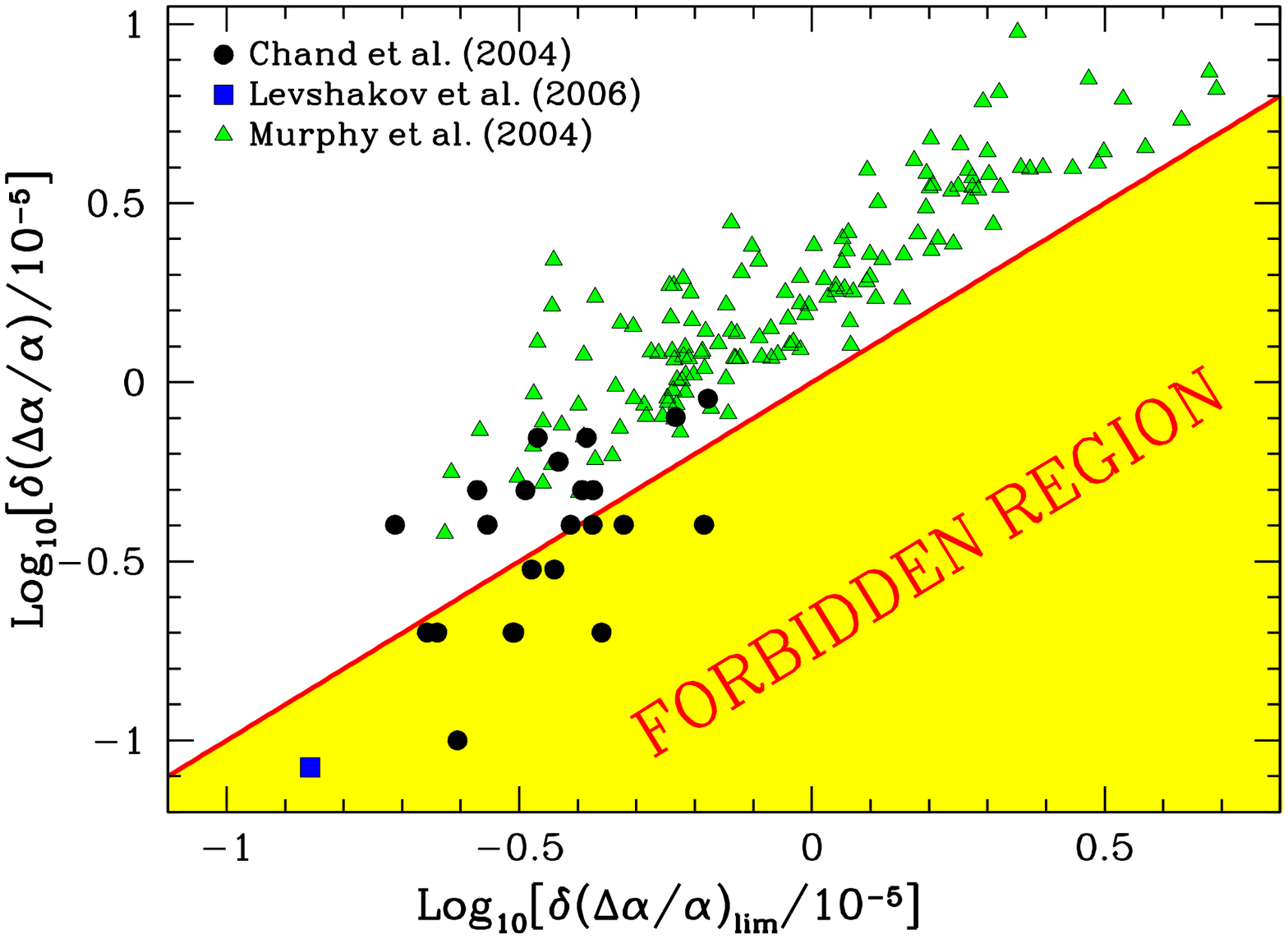}
\vspace{-5mm}
\caption{Quoted errors on $\da$ versus the limiting precision,
  $\delta(\da)_{\rm lim}$, for current samples. The \citet{ChandH_04a}
  and \citet{LevshakovS_06b} samples fail the basic requirement that
  $\delta(\da)$ must be greater than $\delta(\da)_{\rm lim}$.}
\label{fig:lim}
\end{figure}

Figure \ref{fig:lim} shows the 1-$\sigma$ error on $\da$ quoted by the
original authors versus the limiting precision, $\delta(\da)_{\rm
  lim}$. The main results are clear. Firstly, the 1-$\sigma$ errors
quoted for the HIRES sample in \citetalias{MurphyM_04a} always exceed
$\delta(\da)_{\rm lim}$, as expected if the former are robustly
estimated. Secondly, for at least 11 of their 23 absorbers,
\citetalias{ChandH_04a} quote errors which are \emph{smaller} than
$\delta(\da)_{\rm lim}$. Recall that since their error arrays are
larger than ours, 11 out of 23 is a conservative estimate; if they
were to calculate $\delta(\da)_{\rm lim}$ using their larger error
arrays then more points on Fig.~\ref{fig:lim} would shift to the right
into the `forbidden region' where $\delta(\da) < \delta(\da)_{\rm
  lim}$. Finally, the very small error quoted by
\citetalias{LevshakovS_06b} for HE\,0515$-$4414, $0.084\times10^{-5}$,
disagrees significantly with the limiting precision of
$0.14\times10^{-5}$. Thus, the (supposedly) strong current UVES
constraints on $\da$ fail a basic consistency test which not only
challenges the precision reported by \citetalias{ChandH_04a} and
\citetalias{LevshakovS_06b} but which must bring into question the
robustness and validity of their analysis and final $\da$ values.

As discussed in Section \ref{ssec:chi}, underestimated error bars on
$\da$ are expected from the jagged $\chi^2$ curves derived by
\citetalias{ChandH_04a}. Now we explore what effect the fluctuations
had on the values of $\da$ themselves \emph{and} their uncertainties
by applying a robust $\chi^2$ minimization algorithm.

\section{Correcting the analysis of Chand et al.~(2004)}\label{sec:rev}

\subsection{Analysis method}\label{ssec:anal}

\begin{figure*}
\centerline{\hbox{
 \includegraphics[width=0.48\textwidth]{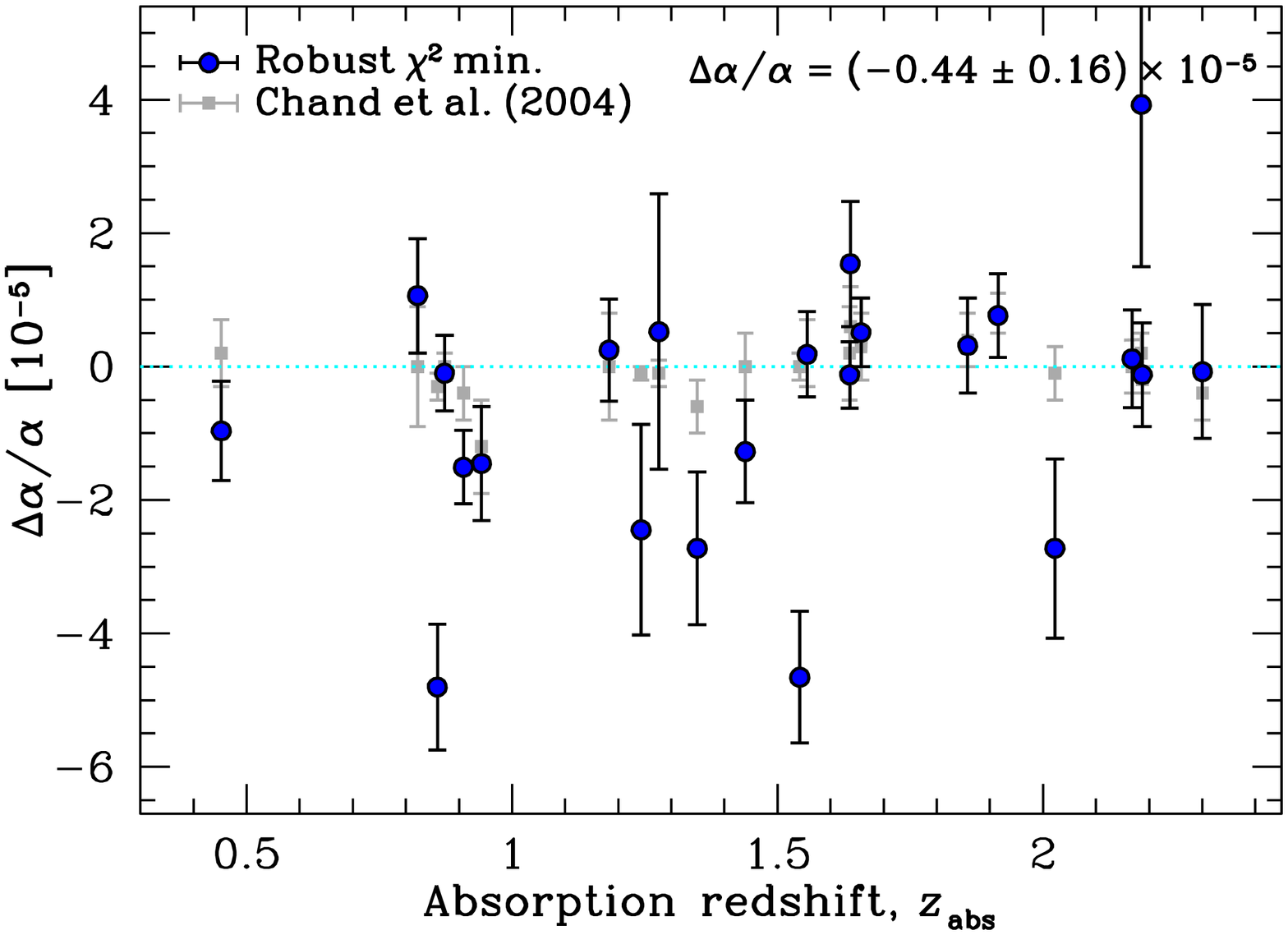}
 \hspace{0.01\textwidth}
 \includegraphics[width=0.48\textwidth]{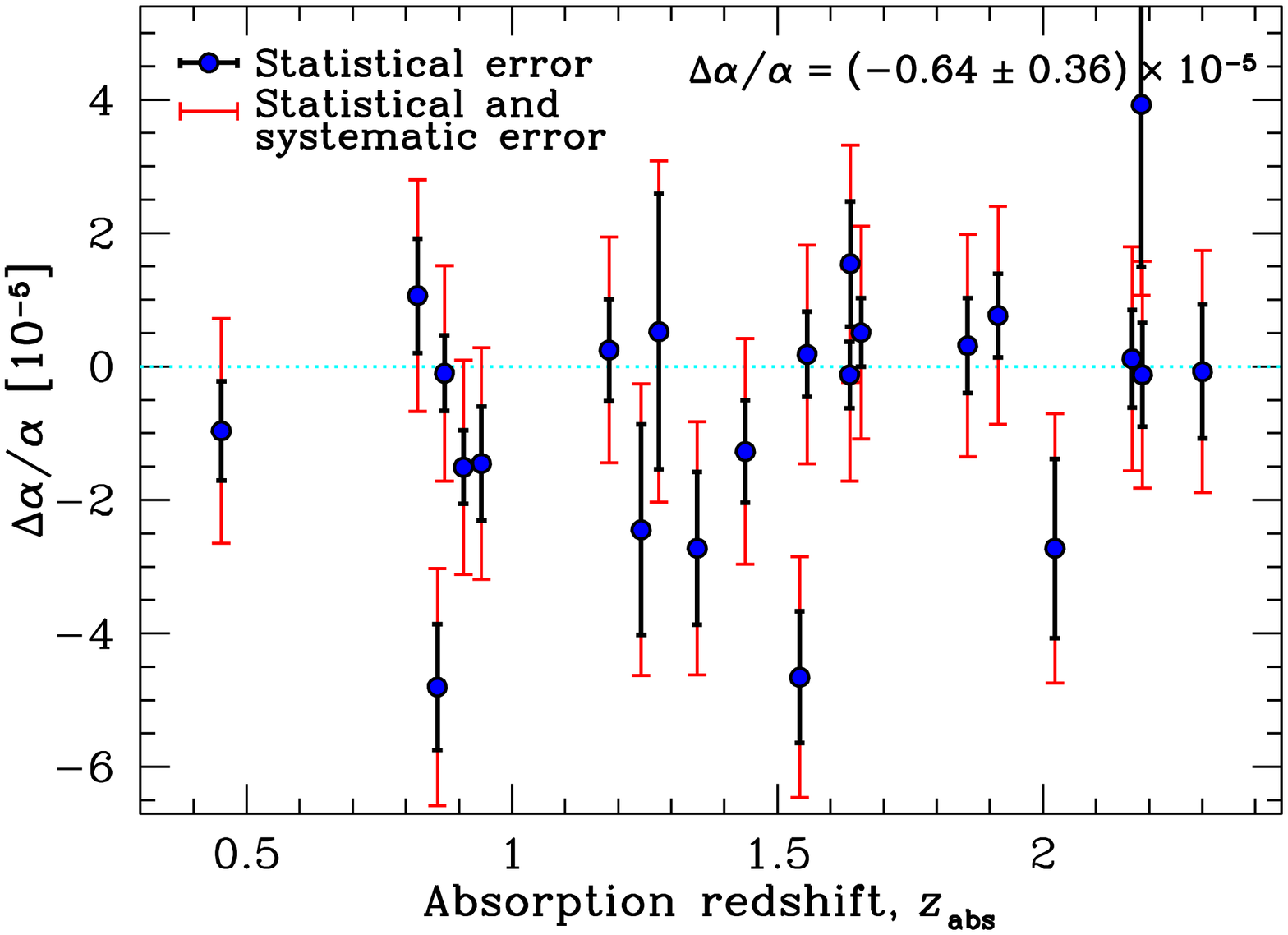}
}}
\vspace{-2mm}
\caption{Left: Our new results (filled circles) are
  inconsistent with those of \citet{ChandH_04a} (grey squares) even
  though the same reduced data and profile fits were used. The only
  difference is that our $\chi^2$ minimization procedure is robust.
  The weighted mean $\da=(-0.44\pm0.16)\times10^{-5}$ is consistent
  with our previous HIRES results but the scatter in the new results
  indicate remaining systematic errors in the data and fits. Right: To
  account for this additional scatter, we increase 1-$\sigma$ errors
  by adding a constant amount in quadrature such that the final
  $\chi^2_\nu=1$ about the weighted mean. The most conservative
  weighted mean result therefore becomes
  $\da=(-0.64\pm0.36)\times10^{-5}$.}
\label{fig:res}
\end{figure*}

It is crucial to emphasize from the beginning here that we use the
same spectra as \citetalias{ChandH_04a} for the following analysis.
To be precise, we use the same values of the flux with precisely the
same wavelength scale, while Section \ref{sssec:apply} details why our
error arrays are somewhat ($\sim$30\,per cent) \emph{smaller} than
those of \citetalias{ChandH_04a}. Our aim is to establish the results
that would have been obtained from the data had the $\chi^2$
minimization algorithm used by \citetalias{ChandH_04a} not failed. For
this reason we also fitted the same velocity structures to the data as
\citetalias{ChandH_04a}. That is, for each absorption system, the
best-fitting Voigt profile parameters of \citetalias{ChandH_04a} were
treated as \emph{first guesses} in our $\chi^2$ minimization
procedure. This is necessary because the Voigt profile parameters they
report are not truly the best fitting ones (because their $\chi^2$
minimization algorithm failed). Nevertheless, the `qualitative'
aspects of the fits -- i.e.~the number and approximate relative
positions of the constituent velocity components -- remained the same
throughout our subsequent $\chi^2$ minimization; it is in this sense
that we state above that our ``velocity structures'' are the same as
those of \citetalias{ChandH_04a}. Finally, the relationships between
the Doppler widths of corresponding velocity components in different
transitions were also the same as in \citetalias{ChandH_04a}.

The analysis procedure was the same as that described in detail in
\citetalias{MurphyM_03a}. The Voigt profile fitting and $\chi^2$
minimization are carried out within {\sc vpfit}, a non-linear
least-squares program designed specifically for analysing quasar
absorption
spectra\footnote{http://www.ast.cam.ac.uk/$\sim$rfc/vpfit.html.},
modified to include a single value of $\da$ as a free parameter for
each absorption system \citep{MurphyM_02b}. The relative tolerance for
halting the $\chi^2$ minimization was set to $\Delta\chi^2_{\rm
  tol}/\chi^2=2\times10^{-7}$ (i.e.~small enough that, even for
profile fits with thousands of degrees of freedom, $\Delta\chi^2$ is
still $\ll1$). Extensive simulations have confirmed the reliability
of this approach (\citealt{MurphyM_02b}; \citetalias{MurphyM_03a}).
The atomic data for the different transitions (i.e.~laboratory
wavelengths, oscillator strengths etc.)  were identical to those used
by \citetalias{ChandH_04a}.

Since $\da$ is a free parameter, its value is determined directly
during the $\chi^2$ minimization of each absorber and its uncertainty,
$\delta(\da)$, is derived from the appropriate diagonal term of the
final covariance matrix. As mentioned above, our error spectra are
significantly smaller (though more appropriate) than those of
\citetalias{ChandH_04a} and so the final $\chi^2$ per degree of
freedom in the fit, $\chi^2_\nu$, is typically $\approx$1.5--4 rather
than $\approx$1 as would be expected if the model fit was appropriate
(see Section \ref{ssec:sys} for further discussion). {\sc vpfit}
therefore increases $\delta(\da)$ by a factor of $\sqrt{\chi^2_\nu}$
and these are the values we report here.  Simulations similar to those
discussed in \citet{MurphyM_02b} and \citetalias{MurphyM_03a} confirm
that such a treatment yields very robust uncertainty estimates (see
also Section \ref{ssec:bias}).

\subsection{Results}\label{ssec:res}

\begin{table*}
\begin{center}
\begin{minipage}{0.835\textwidth}
\caption{Comparison of results from \citet{ChandH_04a} and this paper. Columns
  1 \& 2 give the J2000 and B1950 quasar names; the quasar emission
  redshifts are given in column 3. Column 4 gives the redshifts of the
  absorption systems. Columns 5 \& 6 give the values from
  \citet{ChandH_04a} of $\da$ and $\chi^2$ per degree of freedom,
  $\chi^2_\nu$, for the absorption profile fit. Columns 7 \& 8 give
  the results of our attempt to reproduce those values. Column 9 gives
  our estimate of the wavelength calibration errors derived using the
  method of \citet{MurphyM_07b}. All uncertainty estimates are
  1-$\sigma$.}
\vspace{-2mm}
\label{tab:res}
\begin{tabular}{llcccccccccc}\hline
\multicolumn{2}{c}{Quasar name}                       & $\zem$  & $\zab$   & \multicolumn{2}{c}{\cite{ChandH_04a}}                      & \multicolumn{2}{c}{This work}             & $(\da)_{\rm ThAr}$          \\
\multicolumn{1}{c}{J2000} & \multicolumn{1}{c}{B1950} &         &          & $\da$ [$10^{-5}$]                           & $\chi^2_\nu$ & $\da$ [$10^{-5}$]          & $\chi^2_\nu$ & [$10^{-5}$]                 \\\hline
J000344$-$232355          & HE\,0001$-$2340           & $2.280$ & $0.4524$ & $\phantom{+}0.2\pm0.5$                      & $1.10$       & $          -0.963\pm0.747$ & $3.27$       & $          -0.260\pm0.093$  \\
J000344$-$232355          & HE\,0001$-$2340           & $2.280$ & $2.1854$ & $\phantom{+}0.2\pm0.3$                      & $1.15$       & $\phantom{+}3.926\pm2.431$ & $2.16$       & $\phantom{+}0.145\pm0.099$  \\
J000344$-$232355          & HE\,0001$-$2340           & $2.280$ & $2.1872$ & $          -0.2\pm0.2$                      & $1.20$       & $          -0.122\pm0.774$ & $2.10$       & $          -0.089\pm0.099$  \\
J000448$-$415728          & Q\,0002$-$422             & $2.760$ & $1.5419$ & $\phantom{+}0.0\pm0.2$                      & $0.66$       & $          -4.655\pm0.988$ & $1.00$       & $          -0.090\pm0.103$  \\
J000448$-$415728          & Q\,0002$-$422             & $2.760$ & $2.1679$ & $\phantom{+}0.0\pm0.4$                      & $1.03$       & $\phantom{+}0.115\pm0.731$ & $0.78$       & $          -0.102\pm0.069$  \\
J000448$-$415728          & Q\,0002$-$422             & $2.760$ & $2.3006$ & $          -0.4\pm0.4$                      & $0.99$       & $          -0.075\pm1.001$ & $2.54$       & $          -0.066\pm0.086$  \\
J011143$-$350300          & Q\,0109$-$3518            & $2.410$ & $1.1827$ & $\phantom{+}0.0\pm0.8$                      & $0.98$       & $\phantom{+}0.249\pm0.764$ & $1.70$       & $\phantom{+}0.011\pm0.107$  \\
J011143$-$350300          & Q\,0109$-$3518            & $2.410$ & $1.3489$ & $          -0.6\pm0.4$                      & $1.08$       & $          -2.724\pm1.144$ & $2.28$       & $\phantom{+}0.091\pm0.083$  \\
J012417$-$374423          & Q\,0122$-$380             & $2.189$ & $0.8221$ & $\phantom{+}0.0\pm0.9$                      & $0.87$       & $\phantom{+}1.062\pm0.859$ & $2.27$       & $          -0.088\pm0.078$  \\
J012417$-$374423          & Q\,0122$-$380             & $2.189$ & $0.8593$ & $          -0.3\pm0.2$                      & $1.29$       & $          -4.803\pm0.941$ & $2.81$       & $\phantom{+}0.027\pm0.078$  \\
J012417$-$374423          & Q\,0122$-$380             & $2.189$ & $1.2433$ & $          -0.1\pm0.1$                      & $0.89$       & $          -2.447\pm1.579$ & $4.10$       & $\phantom{+}0.376\pm0.095$  \\
J024008$-$230915          & PKS\,0237$-$23            & $2.223$ & $1.6359$ & $\phantom{+}0.2\pm0.7$                      & $0.82$       & $          -0.124\pm0.498$ & $2.00$       & $          -0.062\pm0.105$  \\
J024008$-$230915          & PKS\,0237$-$23            & $2.223$ & $1.6372$ & $\phantom{+}0.6\pm0.6$                      & $1.16$       & $\phantom{+}1.539\pm0.939$ & $2.93$       & $\phantom{+}0.054\pm0.068$  \\
J024008$-$230915          & PKS\,0237$-$23            & $2.223$ & $1.6574$ & $\phantom{+}0.3\pm0.5$                      & $0.92$       & $\phantom{+}0.510\pm0.514$ & $2.29$       & $\phantom{+}0.121\pm0.127$  \\
J045523$-$421617          & Q\,0453$-$423             & $2.660$ & $0.9084$ & $          -0.4\pm0.4$                      & $1.82$       & $          -1.507\pm0.549$ & $4.21$       & $          -0.141\pm0.122$  \\
J045523$-$421617          & Q\,0453$-$423             & $2.660$ & $1.8584$ & $\phantom{+}0.4\pm0.4$                      & $1.13$       & $\phantom{+}0.315\pm0.712$ & $3.77$       & $\phantom{+}0.467\pm0.118$  \\
J134427$-$103541          & HE\,1341$-$1020           & $2.134$ & $0.8728$ & $\phantom{+}0.0\pm0.2$                      & $1.19$       & $          -0.100\pm0.567$ & $2.49$       & $          -0.065\pm0.071$  \\
J134427$-$103541          & HE\,1341$-$1020           & $2.134$ & $1.2767$ & $          -0.1\pm0.2$                      & $1.01$       & $\phantom{+}0.524\pm2.062$ & $4.30$       & $\phantom{+}0.531\pm0.097$  \\
J134427$-$103541          & HE\,1341$-$1020           & $2.134$ & $1.9154$ & $\phantom{+}0.8\pm0.3$                      & $1.49$       & $\phantom{+}0.767\pm0.627$ & $2.08$       & $\phantom{+}0.058\pm0.072$  \\
J135038$-$251216          & HE\,1347$-$2457           & $2.534$ & $1.4393$ & $\phantom{+}0.0\pm0.5$                      & $1.10$       & $          -1.272\pm0.767$ & $4.60$       & $\phantom{+}0.024\pm0.114$  \\
J212912$-$153841          & PKS\,2126$-$158           & $3.268$ & $2.0225$ & $          -0.1\pm0.4$                      & $1.19$       & $          -2.725\pm1.344$ & $2.65$       & $\phantom{+}0.034\pm0.111$  \\
J222006$-$280323          & HE\,2217$-$2818           & $2.406$ & $0.9425$ & $          -1.2\pm0.7$                      & $0.90$       & $          -1.453\pm0.852$ & $2.43$       & $          -0.258\pm0.114$  \\
J222006$-$280323          & HE\,2217$-$2818           & $2.406$ & $1.5558$ & $\phantom{+}0.2\pm0.5$                      & $1.22$       & $\phantom{+}0.183\pm0.639$ & $2.93$       & $          -0.112\pm0.114$  \\\hline
\end{tabular}
\end{minipage}
\end{center}
\end{table*}

The best-fitting values of $\da$ and the 1-$\sigma$ uncertainties,
$\delta(\da)$, are compared with those of \citetalias{ChandH_04a} in
Fig.~\ref{fig:res}(left). Table \ref{tab:res} provides the numerical
results. Many of the $\da$ values are significantly different to those
of \citetalias{ChandH_04a}, typically deviating from zero by much
larger amounts. Moreover, our uncertainty estimates are almost always
larger, usually by a significant margin (even though, again, our error
spectra are consistently \emph{smaller} than those employed by
\citetalias{ChandH_04a}). The formal weighted mean over the 23
absorbers is
\begin{equation}\label{eq:da_raw}
\da=(-0.44\pm0.16)\times10^{-5}\,.
\end{equation}
At first glance, this indicates a significantly smaller $\alpha$ in
the absorption clouds compared to the laboratory value and agrees well
with our previous results from HIRES. However,
Fig.~\ref{fig:res}(left) also reveals significant scatter in the
results well beyond what is expected based on our estimates of
$\delta(\da)$: the value of $\chi^2$ about the weighted mean is $77.6$
which, for 22 degrees of freedom, has a probability of just
$4\times10^{-8}$ of being larger. It is therefore unclear whether
these new results support our previous HIRES results or not. Further
evidence is provided in Section \ref{ssec:bias} that these results do
indeed reflect the \emph{reduced data and profile fits} of
\citetalias{ChandH_04a} and that the discrepancy with their results is
due to strong biases in both their $\da$ and $\delta(\da)$ values.
Systematic errors likely to affect the spectra are also discussed.

If we regard the large additional scatter in the new results as
evidence of some additional random error then we can estimate its
magnitude by adding a constant amount in quadrature to the current
errors such that the final $\chi^2$ about the weighted mean becomes
unity per degree of freedom. The additional random error required is
$1.51\times10^{-5}$. Figure \ref{fig:res}(right) shows the new results
with error bars which include the additional error term. Using these
increased error bars, the weighted mean becomes
\begin{equation}\label{eq:da_fid}
\da=(-0.64\pm0.36)\times10^{-5}\,.
\end{equation}
This value is the most conservative estimate of $\da$ given the
\emph{reduced data and profile fits} of \citetalias{ChandH_04a} as
inputs. A similar procedure for dealing with increased scatter (beyond
that expected from the purely statistical error bars) was followed in
several previous works (\citealt{WebbJ_99a}; \citetalias{MurphyM_03a};
\citealt{TzanavarisP_05a,TzanavarisP_07a}). In Section \ref{ssec:sys}
we discuss the possible origin of the large additional scatter,
concluding that the profile fits of \citetalias{ChandH_04a} are
probably too simplistic. Indeed, this would lead to an additional
systematic error for each absorption system which is random in sign
and magnitude from absorber to absorber. That is, the assumptions
underlying the derivation of equation (\ref{eq:da_fid}) above are
fairly realistic.

\subsection{Biases in previous results}\label{ssec:bias}

\begin{figure*}
\includegraphics[width=0.8\textwidth]{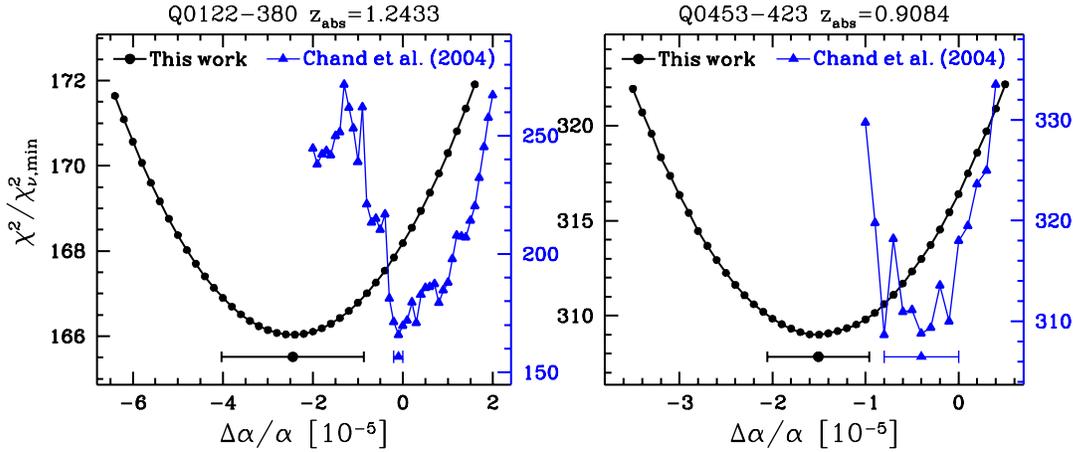}
\vspace{-2mm}
\caption{Example $\chi^2$ curves from our minimization algorithm
  (black circles, left-hand vertical scales) compared with those of
  \citet{ChandH_04a} (grey/blue triangles, right-hand vertical scales)
  which are simply reproduced from their figure 14.  Fluctuations in
  the latter indicate failings in the minimization routine. Also note
  that sometimes (e.g.~right-hand panel) the minima of
  \citetalias{ChandH_04a}'s curves are not well-defined due to these
  fluctuations. The points and error-bars at the minima indicate
  best-fitting values and 1-$\sigma$ uncertainties; for our curves
  they are from the algorithm where $\da$ is a free parameter. The
  vertical scales are normalized by the minimum value of $\chi^2$ per
  degree of freedom (the \citetalias{ChandH_04a} values of
  $\chi^2_\nu$ are reported in their table 3). Thus, the two curves on
  each panel have the same value at their minima since this is simply
  the number of degrees of freedom, $\nu$, by construction. However,
  \citetalias{ChandH_04a}'s curves rise much faster away from $\da=0$
  because the minimization algorithm fails to find the true minimum
  $\chi^2$ at each step in $\da$. This may be the cause of the strong
  evident bias towards zero in \citetalias{ChandH_04a}'s values of
  $\da$ (see Section \ref{ssec:bias}).}
\label{fig:chi}
\end{figure*}

Part of our motivation for revising the analysis of
\citetalias{ChandH_04a} was the large fluctuations in their $\chi^2$
curves. Although we include $\da$ as a free parameter in our $\chi^2$
minimization process, treating it as an external parameter instead (as
in \citetalias{ChandH_04a}) is a simple matter. As discussed in
Section \ref{ssec:chi}, a valid measurement of $\da$ and (especially)
$\delta(\da)$ can \emph{only} come from a smooth, near parabolic,
$\chi^2$ curve. The importance of this point is obvious in
Fig.~\ref{fig:chi} which shows our $\chi^2$ curves in two example
absorbers. For all 23 absorbers, we recover smooth, near parabolic
$\chi^2$ curves, the minima of which coincide well with the values of
$\da$ plotted in Fig.~\ref{fig:res}. The two examples in
Fig.~\ref{fig:chi} are no exceptions. Furthermore, the values of
$\delta(\da)$ recovered from the width of the $\chi^2$ curves near
their minima agree with the values recovered from the covariance
matrix analysis discussed in Section \ref{ssec:anal}. Thus, it is
clear that our $\chi^2$ minimization procedure returns robust values
of $\da$ and $\delta(\da)$.

In contrast, Fig.~\ref{fig:chi} also shows the $\chi^2$ curves of
\citetalias{ChandH_04a}, digitized from their figure 14, for the two
example absorbers. The large $\chi^2$ fluctuations are obvious and it
is clear that they cause two effects: (i) as already discussed, the
1-$\sigma$ uncertainties, $\delta(\da)$, are underestimated and (ii)
the values of $\da$ themselves may be biased towards zero.

The first effect is easy to understand: the minimum $\chi^2$ must, by
definition, be found in a downward extreme fluctuation. Since the
fluctuations are $\ga$1, the width of the `curve' at $\chi^2_{\rm
  min}+1$ will be underestimated, typically by factors of order a few.
The absorber at $z_{\rm abs}=1.2433$ towards Q0122$-$380, shown in
Fig.~\ref{fig:chi}(left), is the extreme example of this problem. The
$\chi^2$ fluctuations are $\sim$10 here, leading
\citetalias{ChandH_04a} to assign an uncertainty of just
$\delta(\da)=0.1\times10^{-5}$ for this system. Note that different
points on their $\chi^2$ curve are separated by this value so, even in
principle, such an error estimate is questionable. Our robust error
estimate is almost a factor of 16 times larger. Also note that it is
far from clear how \citetalias{ChandH_04a} determine $\delta(\da)$ in
some absorbers. One such case is shown in Fig.~\ref{fig:chi}(right)
where, following the previous example, $\da$ would appear to be
$(-0.8\pm0.1)\times10^{-5}$ rather than the value quoted by
\citetalias{ChandH_04a}, $(-0.4\pm0.4)\times10^{-5}$. In a few
absorbers, such as this one, the confusion inherent in deriving
$\delta(\da)$ from such jagged $\chi^2$ curves may have somewhat
ameliorated potentially gross underestimates; in others, not.

\begin{figure}
\includegraphics[width=\columnwidth]{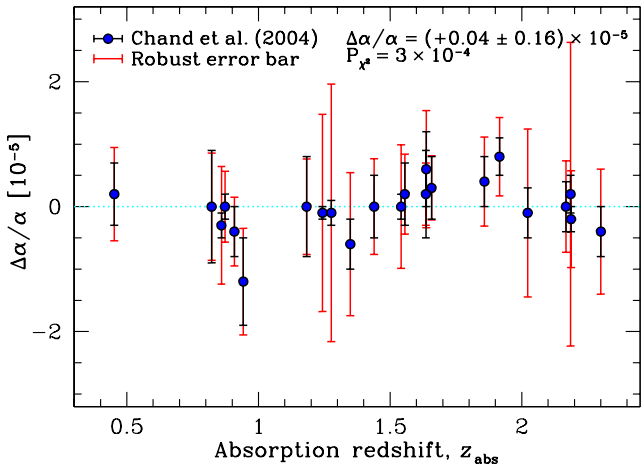}
\vspace{-5mm}
\caption{The 1-$\sigma$ uncertainties from our analysis
  (grey/red bars) compared with those of \citet{ChandH_04a}
  (black bars). The tight distribution of $\da$ values from
  \citet{ChandH_04a} has a probability of just
  $P_{\chi^2}=3\times10^{-4}$ of occurring by chance given our new
  robust error estimates.}
\label{fig:cmp}
\end{figure}

The second effect -- that the $\da$ values of \citetalias{ChandH_04a}
are biased towards zero -- is more difficult to understand. First, to
demonstrate that the effect is significant, Fig.~\ref{fig:cmp} shows
the values of $\da$ from \citetalias{ChandH_04a} plotted with the
1-$\sigma$ uncertainties from our analysis. The $\da$ values are
clearly more tightly clustered around zero than expected based on our
robust error-bars: the value of $\chi^2$ around the weighted mean of
$\da=(0.04\pm0.16)\times10^{-5}$ is just $6.1$. For 22 degrees of
freedom, a $\chi^2$ this low (or lower) \emph{has a probability of
  only $P_{\chi^2}=3\times10^{-4}$ of occurring by chance alone}. The
explanation for such a strong bias \emph{may} be linked, again, to the
failure of the $\chi^2$ minimization algorithm of
\citetalias{ChandH_04a}. One possibility is that, for a given
absorber, the minimization algorithm may have been run several times
with $\da$ fixed to zero with very slightly different initial
conditions (as one might do when experimenting with different velocity
structures in the model fit), thus reducing $\chi^2$ to a relatively
low value even though the algorithm was impaired. When subsequently
using non-zero values of $\da$ in individual minimizations, the
perturbation which the small shift in $\alpha$ imparts to the fit
could cause $\chi^2$ to preferentially fluctuate to higher values. The
impaired minimization algorithm may or may not repair this fluctuation
(i.e.~reduce $\chi^2$ by adjusting the parameters of the fit); this
`hysteresis' would therefore bias $\da$ towards zero.

\subsection{Likely systematic errors}\label{ssec:sys}

Although Fig.~\ref{fig:chi} demonstrates the robustness of our $\da$
and $\delta(\da)$ estimates, the large scatter of the results in
Fig.~\ref{fig:res} is inconsistent with our previous HIRES results.
Furthermore, the HIRES values had a scatter consistent with that
expected from the $\delta(\da)$ estimates (e.g.~\citealt{WebbJ_99a};
\citetalias{MurphyM_03a,MurphyM_04a}). What systematic effects might
contribute to the additional scatter in Fig.~\ref{fig:res}?

We considered a wide variety of systematic effects on $\da$ in
\citet{MurphyM_01b} and \citetalias{MurphyM_03a}, the most obvious
possibility being wavelength calibration errors. The quasar spectra
are wavelength calibrated by comparison with exposures of a
thorium-argon (ThAr) emission-line lamp. A simple test for
miscalibration effects was described in \citet{WebbJ_99a} and applied
to the HIRES data in \citet{MurphyM_01b} and \citetalias{MurphyM_03a}:
the basic approach was to treat the ThAr emission lines near the
redshifted quasar absorption lines to the same MM analysis, thereby
deriving a correction, $(\da)_{\rm ThAr}$, to the value of $\da$ in
each absorber. For the HIRES spectra, wavelength calibration errors
contributed negligible corrections, especially since so many
absorption systems (128) were used \citepalias{MurphyM_03a}.

This ThAr test was not applied to the results of
\citetalias{ChandH_04a}.  However, recently in \citet{MurphyM_07b} we
found that corruptions of the input list of ThAr wavelengths caused
significant distortions of the wavelength scale in UVES spectra such
as those of \citetalias{ChandH_04a}. From these distortions it is
possible to quantify the value of $(\da)_{\rm ThAr}$ and it was
demonstrated in \citet{MurphyM_07b} that the absorber at $z_{\rm
  abs}=1.2433$ towards Q0122$-$380 [Fig.~\ref{fig:chi}(left)] was,
again, particularly problematic, having $(\da)_{\rm
  ThAr}=+0.4\times10^{-5}$. This is 4 times larger than the formal
uncertainty quoted by \citetalias{ChandH_04a} for this system.
However, for most absorbers the corrections due to wavelength
calibration errors are small compared to the scatter in
Fig.~\ref{fig:res}; other systematic errors must dominate.

Another strong possibility is that too few velocity components have
been fitted to the absorption profiles in many of the 23 absorbers.
If this is the case then one should expect to find values of
$\chi^2_\nu$ for the profile fits exceeding $\approx$1, whereas
\citetalias{ChandH_04a} generally found $\chi^2_\nu\sim1$ for their
fits. However, as mentioned several times above, the error spectra
employed by \citetalias{ChandH_04a} were set too high by a factor of
$\approx$1.3--2. This is easily cross-checked by comparing the
r.m.s.~flux in continuum regions around the absorption profiles with
the 1-$\sigma$ error spectra. Thus, when more appropriate error arrays
are employed, as in our new analysis, the velocity structures of
\citetalias{ChandH_04a} prove too simplistic, resulting in the high
values of $\chi^2_\nu\approx1.5$--4 we derive. Clearly -- and, in many
cases, this is obvious simply upon visual inspection -- more velocity
components must be fitted in almost all 23 absorption systems to
account for the evident velocity structure that the high $\chi^2_\nu$
values reflect.  Preliminary fits which include additional velocity
components indicate that `under-fitting' of the absorption profiles
has indeed caused large, spurious excursions in $\da$ and may well be
responsible for the bulk of the scatter in Fig.~\ref{fig:res}.  The
following Section presents a discussion and simulations of complicated
velocity structures which confirm this.

\section{Other constraints from UVES spectra}\label{sec:other}

\begin{figure*}
\centerline{
 \hbox{
  \includegraphics[height=0.42\textwidth,angle=270]{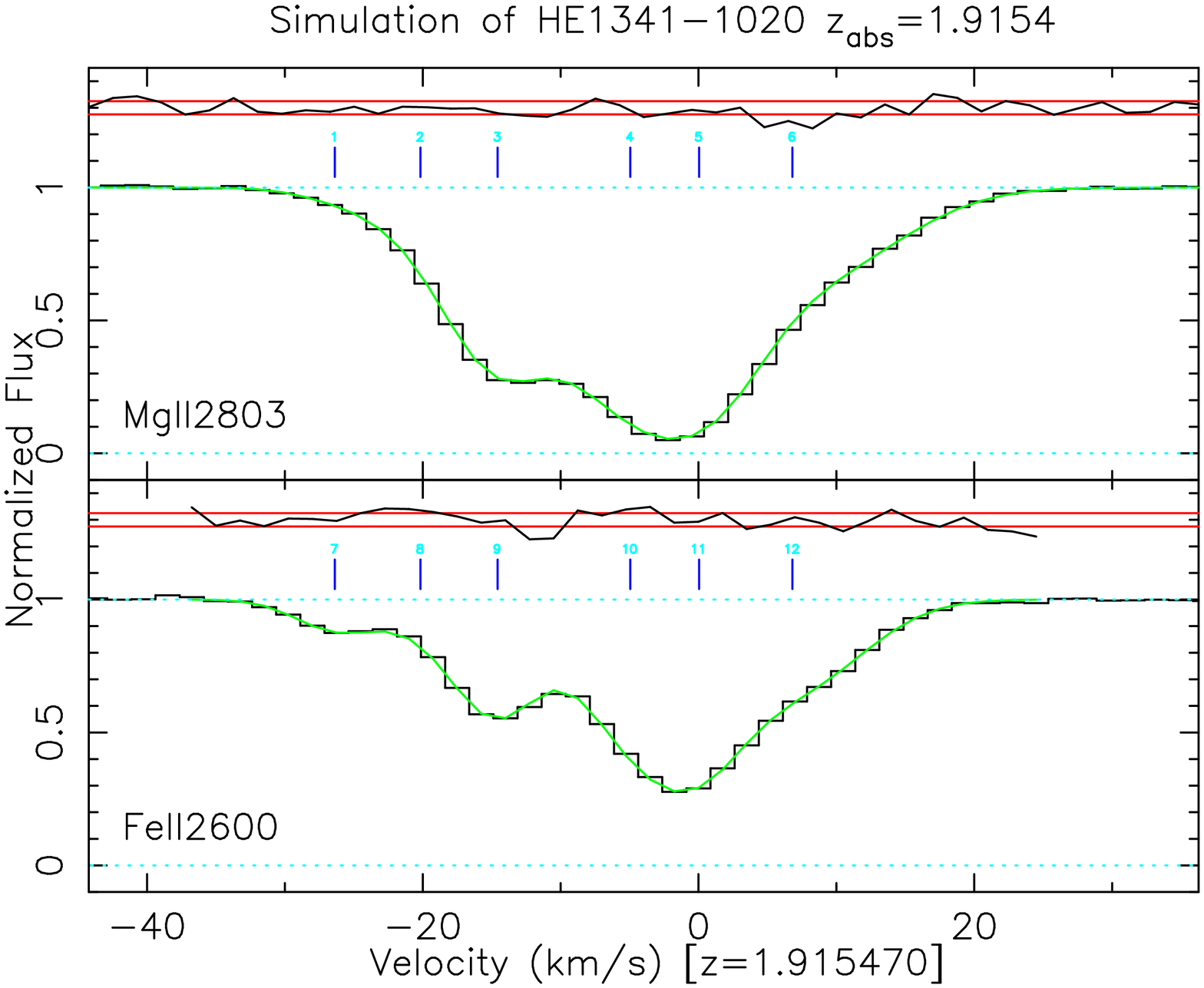}
  \hspace{0.01\textwidth}
  \includegraphics[width=0.49\textwidth]{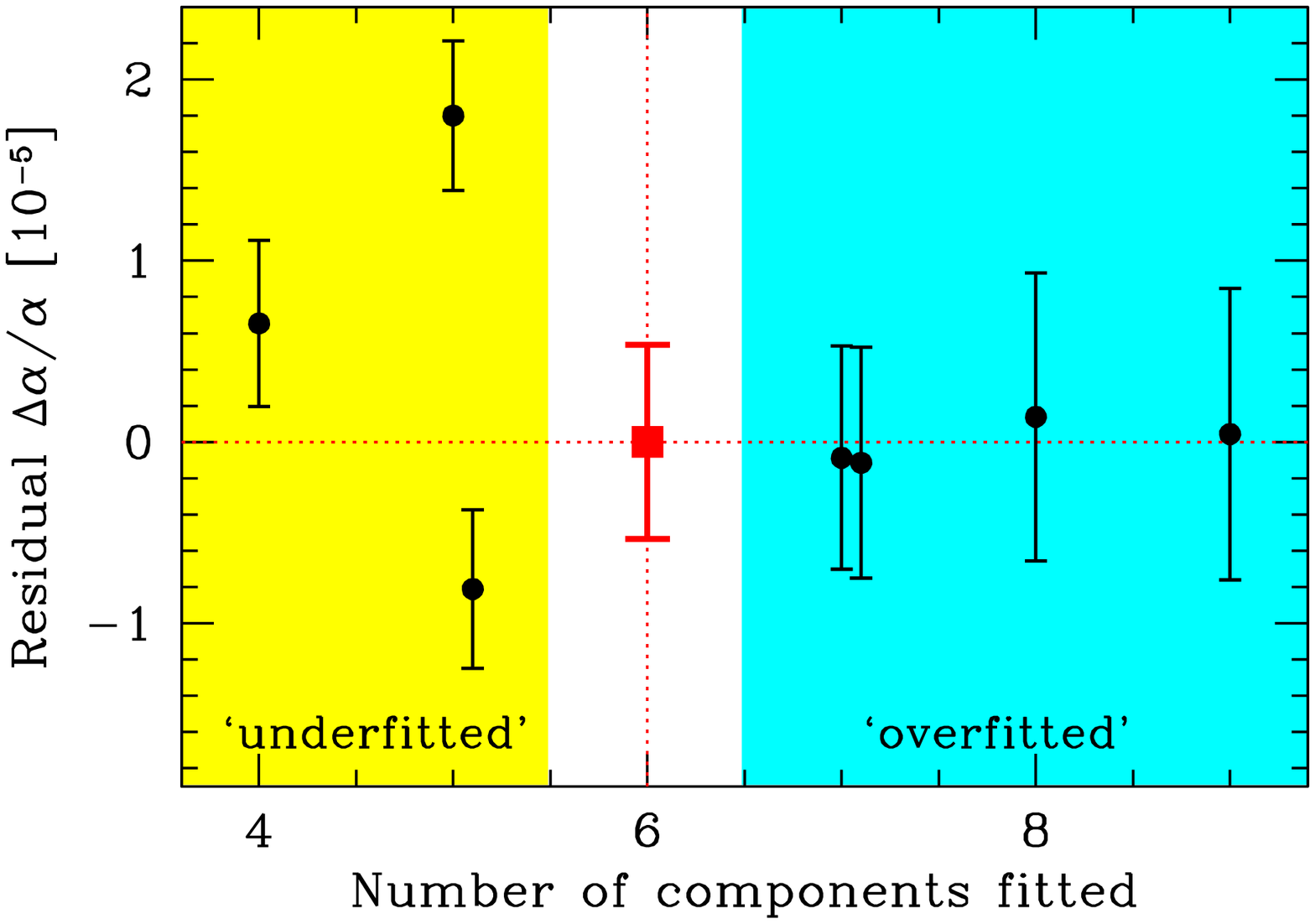}
 }
}
\vspace{-2mm}
\caption{Fitting a simulated absorption profile with different numbers
  of velocity components. Left: Two transitions from the simulation
  showing the 6-component profile (green/grey curve) used to generate
  the synthetic data with $\SNR=200$\,per 2.5-\kms\ pixel (black
  histogram). When under-fitting the data, we removed either or both
  the second and fourth components from the left. When over-fitting we
  added weak components around $-10$ and $10$--$20$\,\kms. Right:
  Values of $\da$ obtained from the $\chi^2$ minimization. For each
  velocity structure (i.e.~each fit with a different number and/or
  placement of components) we plot the mean value of $\da$ and its
  standard deviation over 1000 realizations of the simulated data (the
  mean 1-$\sigma$ uncertainty from the realizations was also very
  close to the standard deviation). Note the large scatter in $\da$
  when the profile is fitted with fewer components than really present
  (`under-fitted') and the small scatter when it is `over-fitted'.}
\label{fig:fit}
\end{figure*}

As mentioned in Section \ref{sec:intro}, several other recent studies
of UVES QSO spectra have ostensibly provided constraints on $\da$. The
first of these was the AD analysis of 15 Si{\sc \,iv} doublets by
\citet{ChandH_05a}. However, the $\chi^2$ curves they present contain
strong fluctuations just like those in \citetalias{ChandH_04a} -- see
their figures 1--6. It is therefore with caution that their final
weighted mean result of $\da=(+0.15\pm0.43)\times10^{-5}$
($1.5<\zab<3.0$) should be interpreted. Our analysis of a sample of 21
somewhat lower \SNR\ Si{\sc \,iv} absorbers from Keck/HIRES gave
$\da=(-0.5\pm1.3)\times10^{-5}$ over the range $2.0<\zab<3.1$ without
similar problems in minimizing $\chi^2$ for each absorber
\citep{MurphyM_01c}.

In Section \ref{sssec:apply} we saw that the MM analysis of the single
complex $\zab=1.151$ absorber towards HE\,0515$-$4414 by
\citetalias{LevshakovS_06b} gave a very small uncertainty of
$\delta(\da)=0.084\times10^{-5}$ but that the limiting precision was
substantially larger, $\delta(\da)_{\rm lim}=0.14\times10^{-5}$.
However, in this case, we cannot easily identify the cause of the
inconsistency. Nevertheless, it is quite possible, even likely, that
the cause of the underestimated uncertainty also affected the value of
$\da$ and, again, caution should evidently be used in interpreting the
result of \citetalias{LevshakovS_06b}. Previous analyses of the same
absorber by the same group utilized very simplistic profile fits: see
figure 2 of \citet{QuastR_04a}, particularly at velocities around
$-20$, $15$, $40$, $50$--$55$, $65$, $85\,\kms$ where large residuals
are clearly visible. As we demonstrate below, these are very likely to
have caused large systematic effects in this single-absorber estimate
of $\da$.

The same single absorber was studied, again with MM analysis, by
\citet{ChandH_06a}. However, in this case the fluctuations on the
$\chi^2$ curve were so large -- $\sim$50; see their figure 9 -- that
the authors found it difficult to define a minimum in the curve.
Instead they attempted to fit a low-order polynomial through the large
fluctuations -- as one would fit a line through noisy data -- to
define a minimum. It must be strongly emphasized that such a practice
is illogical and yields completely meaningless values of $\da$ and
$\delta(\da)$. Firstly, it is not `noise' in the usual sense that one
is attempting to fit through but spurious fluctuations in $\chi^2$
caused by the failure of the algorithm to find the true minimum at
each fixed input value of $\da$. Secondly, the \emph{real} $\chi^2$
curve must lie \emph{beneath} the majority of points on the $\chi^2$
and it \emph{cannot} lie above any of them. The logical conclusion is
that one cannot infer the shape, minimum or width -- i.e.~$\da$ or
$\delta(\da)$ -- from such a $\chi^2$ curve since the real $\chi^2$
curve may lie anywhere below it.

The most recent constraint on $\da$ from UVES QSO spectra was derived
from MM analysis of the $\zab=1.839$ absorber towards Q\,1101$-$264 by
\citetalias{LevshakovS_07a}: $\da=(+0.54\pm0.25)\times10^{-5}$. Three
Fe{\sc \,ii} transitions were employed -- $\lambda$1608, $\lambda$2382
and $\lambda$2600. The latter two have nearly identical
$q$-coefficients so they shift in concert as $\alpha$ varies. However,
the bluer line, $\lambda$1608, shifts in the opposite sense to the
other two and is therefore crucial for any meaningful constraint on
$\da$ to be derived. It is also the weaker line of the trio, with an
oscillator strength less than a third of the others. As outlined in
Section \ref{ssec:sys}, systematic effects, even with small
statistical samples of absorbers, can result if one does not fit the
observed structure in the absorption profiles with an adequate number
of velocity components. For any analysis of a \emph{single} absorber,
this becomes particularly important.

\citetalias{LevshakovS_07a} explored this effect by using three different
fits: a fiducial one with 16 velocity components and two others with
\emph{fewer} (11 and 10) components. They find very similar values of
$\da$ from all fits and therefore maintain that $\da$ is insensitive
to the number of fitted components. However, the components removed
from the fiducial fit to form the 11- and 10-component fits appear
mainly at the edges of the absorption complex and, crucially, do not
absorb statistically significant fractions of the continuum in the
weaker Fe{\sc \,ii} $\lambda$1608 transition. That is, those
components do not appear in Fe{\sc \,ii} $\lambda$1608. Since this
transition is vital to provide any sensitivity to $\alpha$ at all, one
\emph{must} find similar values of $\da$ in all three fits; as far as
$\da$ is concerned, the three fits are identical and the test, in this
case, is ineffective. In any case, visual inspection of the residuals
between the fiducial fit and the data \citepalias[figure 1
in][]{LevshakovS_07a} reveals correlations over $\sim$20-pixel ranges
and, moreover, these ranges occur at similar velocities in the three
different transitions.  This is one indication of \emph{more} than 16
components being required in the fit, not fewer.  We will present
further analysis of this absorber in Bainbridge et al.~(in
preparation).

To illustrate the effect of `under-fitting' absorbers in this way,
Fig.~\ref{fig:fit} shows a simulated absorption spectrum generated
with 6 velocity components but which has then been fitted with
different numbers of components. The Mg{\sc \,ii} doublet
($\lambda\lambda$2796/2803) and the five strongest Fe{\sc \,ii}
transitions longward of 2340\,\AA\ were included in the fit but
Fig.~\ref{fig:fit} just shows two representative transitions. All
transitions were simulated with a \SNR\ of 200 per 2.5-\kms\ pixel in
the continuum. The velocity structure was inspired by (but is not
strictly the same as) a real absorption system -- one of the 23
contained in the \citetalias{ChandH_04a} sample -- and includes two
weaker components (the second and fourth from the left). It is these
components which are missing in our 4- and 5-component fits to the
simulated data.  These fits result in values of $\da$ which deviate
significantly from the input value depending on which component is
removed. Thus, `under-fitting' individual absorption systems can
easily cause spurious values of $\da$.

With such high \SNR\ `data' we can also `over-fit' the simulated
absorption system. The additional components were placed at around
$-10$ and $10$--$20$\,\kms. Of course, $\chi^2_\nu$ for these fits is
below unity and, more importantly, the additional components are not
statistically justified: the reduction in $\chi^2$ they provide
compared to the fiducial 6-component fit is smaller than the
additional number of free parameters they introduce. Nevertheless, the
results show that the value of $\da$ is robust to the introduction of
these components. We hasten to add that one cannot simply keep
fitting additional components: note that the uncertainty on $\da$
increases as one adds components -- with more components, $\chi^2$
becomes increasingly insensitive to $\da$. However, when reporting
strong constraints on $\da$ from \emph{individual} absorbers, some
demonstration of how robust $\da$ is to both under- and over-fitting
is desirable and our illustration in Fig.~\ref{fig:fit} suggests that
the latter is more conservative than the former.

\section{Conclusion}\label{sec:conc}

We have critically analysed the reliability of the MM treatment of 23
VLT/UVES absorption systems by
\citet[\citetalias{ChandH_04a}]{ChandH_04a}. Using the \emph{same data
  and profile fits} we find values of $\da$ in individual absorbers
which deviate significantly from those of \citetalias{ChandH_04a} and
our uncertainty estimates are consistently and, in some cases
dramatically, larger. Indeed, simple (but robust) calculations of the
limiting precision available on $\da$ in these absorbers indicates
that $\ga$\,half of \citetalias{ChandH_04a}'s quoted uncertainties are
impossibly low; the \SNR\ of the spectra and the complexity of the
fitted velocity structures simply do not allow such small
uncertainties on $\da$.

This is altogether unsurprising given the large fluctuations in the
$\chi^2$ curves presented by \citetalias{ChandH_04a}. The only way
such large fluctuations can occur is if the $\chi^2$ minimization
algorithm fails to reach a truly minimum value at each step along the
input $\da$ axis. That is, since the very means by which $\da$ is
estimated (the $\chi^2$ curve) is flawed, so are the values of $\da$
and their uncertainties invalid. Just as the 1-$\sigma$ errors were
underestimated, we have also demonstrated that
\citetalias{ChandH_04a}'s values of $\da$ are strongly biased towards
zero.  Again, this is likely to stem from the fluctuations on the
$\chi^2$ curves.

It is therefore expected that \citetalias{ChandH_04a}'s weighted mean
result of $\da=(-0.06\pm0.06)\times10^{-5}$ should not truly represent
the \emph{reduced data and profile fits} they employed. Indeed, our
own analysis -- again, with the same reduced data and profile fits --
yields a very different central value and much a larger error bar:
$\da=(-0.44\pm0.16)\times10^{-5}$. Since our $\chi^2$ minimization is
demonstrably robust, we argue that this latter value does truly
reflect the reduced data and profile fits. However, we do not argue
that this value is necessarily the final or best one to be gleaned
from this dataset: improvements in the profile fits are almost
certainly required to reduce the evident scatter in the 23 values of
$\da$ to within that expected based on their (robust) 1-$\sigma$
uncertainties. Although not discussed in this paper, improvements in
the data reduction process may also be needed. After increasing the
uncertainties by adding a constant amount in quadrature to match the
scatter, we obtain a more conservative estimate of the weighted mean
from \citetalias{ChandH_04a}'s reduced data and profile fits:
$\da=(-0.64\pm0.36)\times10^{-5}$. Note that the uncertainty here is 6
times larger than that quoted by \citetalias{ChandH_04a}. The data and
fits of \citetalias{ChandH_04a} thus provide no stringent test of the
Keck/HIRES evidence for a varying $\alpha$.

Jagged $\chi^2$ curves have also affected the AD analysis of Si{\sc
  \,iv} doublets in UVES spectra by \citet{ChandH_05a} and the MM
analysis of a single, high-\SNR\ UVES spectrum by \citet{ChandH_06a}.
The same single absorber was studied by \citet{LevshakovS_06b} but
their quoted uncertainty is much lower than the limiting precision
available, even in principle, from this spectrum. The only other UVES
constraint on $\da$ is from another single absorption system studied
by \citet[][\citetalias{LevshakovS_07a}]{LevshakovS_07a}. However, we
have demonstrated that when analysing individual systems in this way,
care must be taken to ensure that all the structure in the absorption
profiles is adequately fitted. Indeed, simulations indicate
`under-fitting' the profile can cause dramatic spurious excursions
from the real value of $\da$ whereas `over-fitting' the profile (to a
mild degree) is a more conservative approach. We argue that additional
components are probably required to fit the data of
\citetalias{LevshakovS_07a}; this will be demonstrated in Bainbridge
et al.~(in preparation).

In summary, reliable comparison of HIRES and UVES constraints on a
varying $\alpha$ must await improvements in the analysis of UVES
spectra. We are currently reducing $>$100 UVES spectra to arrive at an
internally and statistically robust estimate of $\da$ for this
purpose.

\section*{Acknowledgments}

MTM thanks STFC (formerly PPARC) for an Advanced Fellowship at the
IoA. We thank the referee, Simon Morris, for valuable comments which
improved the presentation of the paper.


\bspsmall

\label{lastpage}

\end{document}